# System Analysis and Pre-Flight Evaluation of Deployable Solar Panels for 3U CubeSat HOKUSHIN-1

By Yuji Sakamoto[1], Masaki Aoi[2], Sho Suzuki[2], Takumi Haga[2], Shumpei Hosokawa[2], Yuma Abe[2], Yuya Tasaki[2], Tsuyoshi Totani[2], Sou Nakamura[3], Masaharu Uchiumi[3], Shinya Fujita[1]

[1] *Green Goals Initiative, Tohoku University, Sendai, Japan*
[2] *Division of Mechanical and Aerospace Engineering, Hokkaido University, Sapporo, Japan*
[3] *Aerospace Plane Research Center, Muroran Institute of Technology, Muroran, Japan*

This paper describes the system design methodology derived from the development and evaluation tests of deployable solar panels to be mounted on a 3U CubeSat. The study mainly includes structural analysis, thermal analysis, and a review of vibration test results. Hokkaido University is developing the 3U CubeSat HOKUSHIN-1 in collaboration with Tohoku University and Muroran Institute of Technology. Deployable solar panels are a key technology for future planned lunar exploration missions, as they enable power-intensive communication and propulsion required for orbit control. The satellite also demonstrates a newly developed compact and efficient propulsion system. The satellite has dimensions of approximately 10x10x34 cm, a mass of 3.99 kg, and will be deployed into a circular orbit at an altitude of about 400 km with an orbital inclination of 51.6 degrees from the International Space Station.

## 1. Introduction

Hokkaido University is developing the 3U CubeSat HOKUSHIN-1 in collaboration with Tohoku University and Muroran Institute of Technology. This satellite is equipped with deployable solar panels and will operate in low Earth orbit. This mechanism is essential for realizing long-distance communications and propulsion-based orbit control, both of which require significant electrical power in a planned follow-up lunar exploration mission. In this paper, we present findings from the initial design phase, including material selection and dimensional design, from both structural and thermal perspectives. In addition, the results of vibration tests conducted on the thermal-structural test model are reported. The launch of this satellite is scheduled for after 2024.

The overarching objective of this project is to acquire development and operational technologies for deep-space satellites that can be realized using the limited facilities, personnel, and development budgets typical of universities, with the aim of conducting future missions beyond the Moon. To achieve this objective, technologies such as "orbit determination using a university-operated ground station," "compact and efficient propulsion devices," and "deployable solar panels capable of high power generation" are indispensable. The satellite bus system inherits the design of the ASTERISC satellite, which was jointly developed by Chiba Institute of Technology and Tohoku University and was successfully demonstrated in orbit (Ref. [1]).

In recent years, while referring to commercially available deployable solar panels for CubeSats (EXA DMSA) and recent 3U CubeSats such as CLICK and CPOD (Ref. [2]), a new deployable solar panel design has been developed. This design includes a hold-and-release mechanism and deployable hinges. Achieving an appropriate balance between thermal and structural design is critical. The objectives include weight reduction, minimizing temperature fluctuations between high- and low-temperature conditions, and improving solar cell efficiency. During the operational phase, the satellite performs not only free rotation in a standby attitude but also temporary attitude control for pointing toward the Earth and the Sun. Accordingly, a wide range of thermal environments must be considered, ranging from maximum-eclipse conditions—where the solar incidence angle β with respect to the orbital plane is zero—to full-sun conditions in which continuous solar illumination persists for several days.

Hokkaido University has access to a public vibration test facility located within walking distance of the campus. This facility has previously been used for vibration testing of CubeSat structural test models as well as a 50 cm-class engineering model microsatellite. In addition, the Space Mission Center at Hokkaido University is equipped with a thermal vacuum chamber and has experience conducting environmental tests for 50 cm-class flight-model microsatellites.

## 2. Overview of the Satellite System

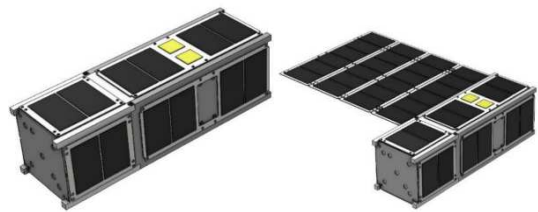

Fig. 1.   Appearance of the HOKUSHIN-1 satellite

Table 1 summarizes the specifications of the satellite system. The satellite has approximate dimensions of 10x10x34 cm and a launch mass of 3.99 kg. Operations will commence from a circular orbit at an altitude of about 400 km with an inclination of 51.6 degrees, following deployment from the International Space Station (ISS). Based on the orbital lifetimes of previously launched CubeSats of similar size, the satellite is expected to re-enter the atmosphere after approximately 16 months. The onboard sensors provide coarse attitude determination capability, primarily measuring the Sun direction, geomagnetic field vector, and coarse angular velocity. A star tracker is not installed. Attitude control is implemented without reaction wheels and relies on external torque generation using magnetic torquers and thrusters. The expected attitude control accuracy is approximately 2–3 degrees, and an orbital altitude control experiment using propulsion is planned.

The power conversion efficiency of the solar cells exceeds 30%. After deployment of the solar array panels, five parallel strings can be oriented toward the Sun, enabling an expected maximum power generation of 34.2 W, compared with a maximum power consumption of 16 W.

The design requirements for the safety review follow the JAXA handbook (document number JX-ESPC-101132-D). For structural stiffness and strength, the first natural frequency is required to exceed 60 Hz. Static load analysis assumes a 9 G launch environment and a compressive load of 46.6 N applied at each rail end, with allowable stress limited to less than 30% of the material tensile strength. In this satellite, A6061P-T651 aluminum alloy is selected as the primary structural material, resulting in an allowable stress of 88.5 MPa based on a minimum tensile strength of 295 MPa (JIS H4000-2017).

The temperature environment prior to deployment is specified as −15 to +60 °C, taking into account the launch phase as well as the internal and external environments of the ISS before release. During on-orbit operation, the temperature of internal equipment is targeted to be maintained within +5 to +40 °C in order to ensure battery health. In addition, the temperature of the main body is required to remain below +50 °C, even under temporary full-sun conditions.

Regarding the solar cells, lower temperatures are favorable because power generation efficiency decreases at elevated temperatures. However, integrated circuit components, such as temperature sensors mounted on the deployable panels, must not be exposed to temperatures outside the typical industrial-grade range of −40 to +80 °C.

As shown in Fig. 2, the satellite consists of a bus module (MOD2), equivalent to a 2U CubeSat, and a propulsion module (MOD1), equivalent to a 1U CubeSat. The two modules are connected using four M4 bolts. The outer structure of MOD1 forms a sealed container housing control boards, piping, valves, and propellant. The criteria for classifying this structure as a closed vessel rather than a pressure system are that the internal energy is less than 19,310 J and the differential pressure relative to the external environment is less than 1.5 atm. The propellant pressure is designed to remain below 1.1 atm, even at the maximum temperature of +60 °C prior to deployment, and stress analysis confirms that vapor pressure loads remain within allowable limits. Since structural damage is unacceptable even after deployment, the maximum operational temperature of MOD1 must be maintained below +60 °C.

Table 1. Specifications of the HOKUSHIN-1 satellite

| | |
|---|---|
| **Size and Weight** | |
| at Launch | W 100 x D 100 x H 340.5 mm, 3.99kg |
| **Orbit** | |
| height | Circle, 400 km, 92.6-min period |
| lifetime | 16 months to orbital decay |
| inclination | 51.6 deg |
| **Attitude Determination and Control** | |
| sensors | MEMS gyro sensor, magnetometers, sun sensors, GPS receiver |
| actuators | 3-axis, magnetic torquers, thrusters |
| **Power** | |
| solar cells | Triple junction cell (30.7% efficiency) 2.41 V x 473mA = 1.14W/pc @peak 6 series x 7 parallels = 42 pcs = 34.2 W (with dep. paddles) |
| batteries | NiMH, 8 series x 1 parallel = 8 cells 9.6V discharge x 2.0Ah = 19.2Wh (total) |
| P. consumption | 1.2 (power save) - 16.0W (max at mission) |
| **Communication** | |
| location | Sendai station, Japan with 2.4-m dish |
| uplink | S-BAND, 1 kbps |
| downlink | S-BAND, 0.1W (100kbps), 0.5W (2Mbps) |

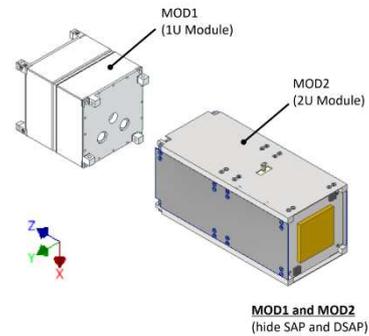

Fig. 2. Bus module (MOD2) and propulsion module (MOD1)

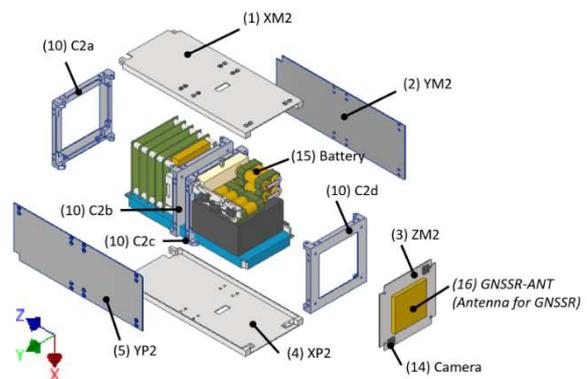

Fig. 3. Structural elements of the bus module (MOD2)

Figure 3 illustrates the structural elements of MOD2. With the exception of the +Z face interfacing with MOD1, the outer panels (Panels 1–5) are fabricated from flat or machined aluminum plates. The components labeled C2a through C2d in part (10) are rectangular frames with dimensions of 10x10 cm. These frames enhance stiffness against static loads in the X and Y directions, where central deflection tends to increase.

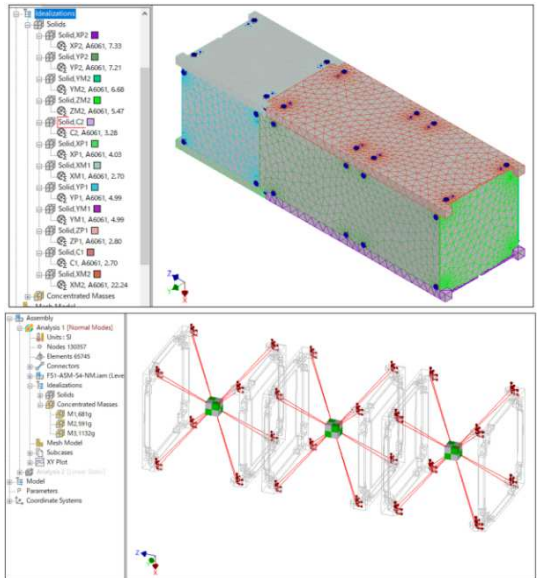

Fig. 4. Stiffness and static load analysis model

Table 2. Equivalent mass density of MOD2 components

| panel/frame | m_str (g) | rho_str (g/cm^3) | m_(str+compo) (g) | rho_(str+compo) (g/cm^3) |
|---|---|---|---|---|
| XP2 | 77 | 2.70 | 209 | 7.33 |
| XM2 | 77 | 2.70 | 635 | 22.24 |
| YP2 | 51 | 2.70 | 135 | 7.21 |
| YM2 | 51 | 2.70 | 125 | 6.68 |
| ZM2 | 46 | 2.70 | 93 | 5.47 |
| C2a | 34 | 2.70 | 42 | 3.28 |
| C2b | 34 | 2.70 | 42 | 3.28 |
| C2c | 34 | 2.70 | 42 | 3.28 |

Internal equipment is mounted to these rectangular frames using bolted connections.

## 3. Structure Analysis for Launch Configuration

Figure 4 shows the stiffness and static load analysis model used at the Preliminary Design Review (PDR) phase. Structural analysis of the satellite was performed using Autodesk Inventor Nastran 2020®. Components other than the primary structure were modeled as lumped mass elements. Rectangular frames with dimensions of 10x10 cm were explicitly modeled, and onboard equipment was rigidly fixed to each frame as concentrated masses. Smaller components and solar panels mounted on the outer surfaces, including deployable panels, were represented as distributed masses applied to the flat panels, with only the material density adjusted accordingly.

In this model, the density of aluminum was set to 2.7 g/cm³, while XP2 and XM2 were modeled with equivalent densities of 7.33 g/cm³ (mass 209 g including fasteners) and 22.24 g/cm³ (mass 635 g), respectively. Panels were converted into solid elements, and bolts were defined as connectors and internally interpreted as beam elements. M3 bolts were used to fasten the main structural components.

As a result of the analysis, the first and second natural frequencies were found to be 232 Hz (effective mass = 0.68

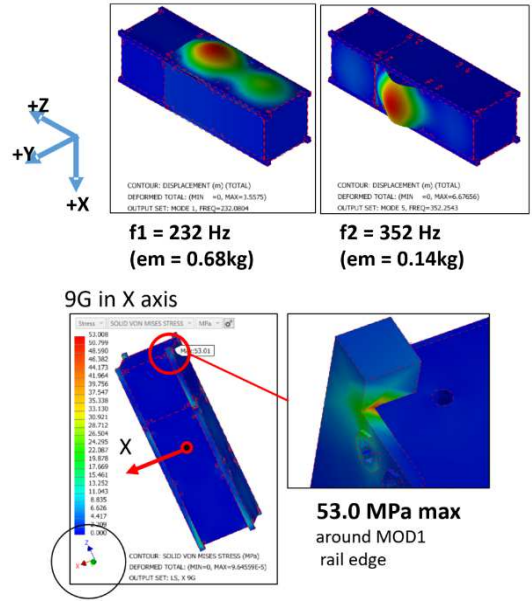

Fig. 5. Results of natural frequency and static load analysis

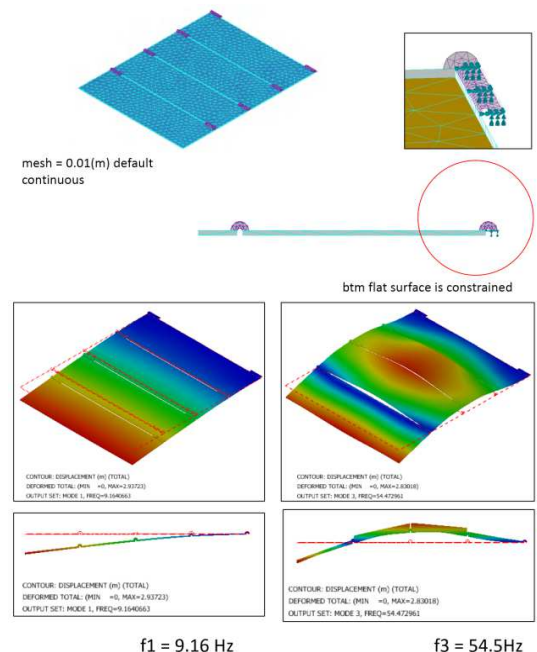

Fig. 6. Structural analysis of the simplified DSAP model (1.2mm thick aluminum: 9.16Hz, 54.5Hz, FR4: 6.11Hz, 36.2Hz)

kg) and 352 Hz (effective mass = 0.14 kg), respectively. The maximum static stress was 53.0 MPa under a 9 G load applied along the X-axis. For all axes under a 9 G load condition, the stress levels remained below the allowable stress limit of 88.5 MPa.

## 4. Structure Analysis of Deployable Solar Panels (DSAP)

After deployment in orbit, insufficient structural stiffness of the deployable panels can lead to increased disturbance during attitude control. Conversely, achievable stiffness is constrained by limitations on dimensions and mass distribution. In this study, the deployable panels were assumed to be fabricated from either FR4 material or aluminum plates. FR4 is a glass-epoxy material commonly used for electronic circuit boards, enabling minimization of physical wiring within the panels. The Young's modulus of aluminum is 68.9 GPa, whereas that of FR4 (e.g., FR402) is 24.1 GPa. A lower Young's modulus results in larger deflections under launch vibration loads, increasing the risk of exceeding the allowable envelope, which extends 6.5 mm beyond the CubeSat rail surface planes.

As illustrated in Fig. 6, a simplified structural model of the DSAP was defined. Based on mass allocation, 468 g was assigned to a single panel assembly (117 g per sheet), including solar cells, adhesive layers, wiring, and structural panels. Drawing on results from commercially available products, two spring hinges were installed at both ends of each panel. Considering the thickness of the solar cells and adhesive layers, the panel thickness was set to 1.2 mm for both aluminum and FR4 options, in accordance with the 6.5 mm envelope constraint outside the CubeSat rails. A 2 mm gap between adjacent panels was introduced to avoid interference between glass-covered solar cells. The simplified hinge model had an inner diameter of 2 mm, an outer diameter of 6 mm, and a length of 24 mm. Due to its small size, the hinge case was machined from aluminum and combined with a stainless-steel hinge pin and a custom torsion spring.

For the finite element analysis, a standard mesh size of 10 mm was adopted. The boundary conditions were defined by constraining two planar surfaces at the hinge locations on the edge panels. As a result, the natural frequencies for the 1.2 mm aluminum panel configuration were 9.16 Hz and 54.5 Hz. As a reference, the maximum deflection under a 1 G load was 4.7 mm, and the allowable acceleration derived from the allowable stress of 88.5 MPa (30% of the tensile strength) was calculated to be 3.14 G in the direction normal to the panel surface. Although a 1 G environment does not occur during on-orbit operation, such loading conditions may arise during ground handling and testing.

For the 1.2 mm FR4 panel configuration, the natural frequencies were 6.11 Hz and 36.2 Hz, and the acceleration corresponding to the allowable stress limit was 2.97 G.

## 5. Thermal Analysis after DSAP Deployment

The on-orbit temperatures of individual panels after deployment of the deployable solar array panels (DSAP) are estimated. Because the surface materials of the lightweight panels have a strong influence on temperature variation, surface conditions must be selected carefully. The candidate surface materials considered in this study include bare aluminum, heat-control aluminum tape, FR4 substrate, and polyimide tape. The thermal analysis in this section is based on a single-node model, in which radiative heat transfer between the satellite and the external environment is considered, while conductive heat transfer within the satellite is completely neglected. The satellite parameters used in the analysis are the mass, specific heat, and surface areas—two panels for the DSAP case and six panels for the satellite main body. A self-developed analysis code incorporating orbital motion and attitude dynamics is used.

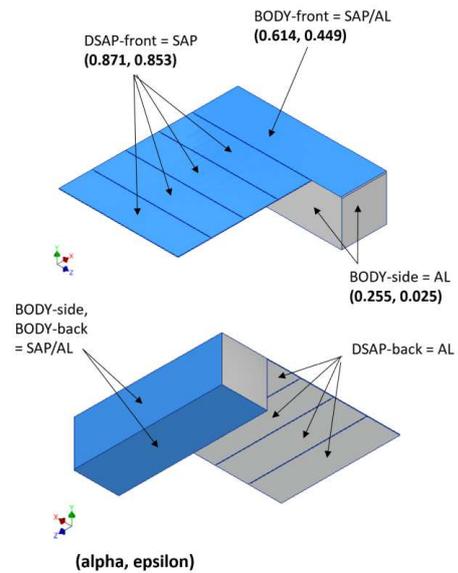

Fig. 7. Model example for thermal analysis and thermo-optical properties of each surface

Table 3. Dimensions, mass and thermal properties of elements

|  | dimension | mass | heat spec. C = J/(kg.K) K = W/(m.K) |
|---|---|---|---|
| BODY (A6061) | 327x100 x100 mm | 3518g (1.076 g/cm^3) | C = 897 K = 167 |
| DSAP (FR4) | 261 x 73 x 1.2 mm (each) | 117g (5.165 g/cm^3) x 4pcs | C = 950 K = 0.44 |
| total |  | 3986g |  |

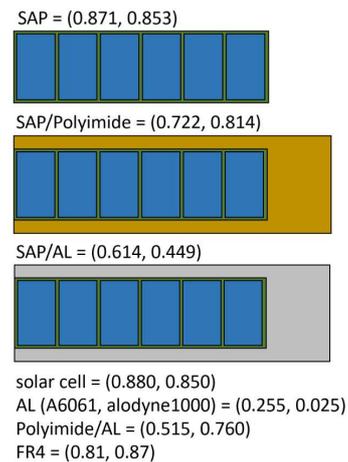

SAP = (0.871, 0.853)
SAP/Polyimide = (0.722, 0.814)
SAP/AL = (0.614, 0.449)

solar cell = (0.880, 0.850)
AL (A6061, alodyne1000) = (0.255, 0.025)
Polyimide/AL = (0.515, 0.760)
FR4 = (0.81, 0.87)

Fig. 8. Solar cell arrangement on each surface and thermo-optical properties of materials (α, ε)

The temperature variation in the single-node model is governed by an energy balance involving multiple heat flux terms. The detailed formulation is omitted here, as it is described extensively in standard satellite thermal analysis handbooks. In this model, $Q_i$ represents individual heat flux components, including solar heat input ($Q_S$), Earth infrared radiation ($Q_I$), albedo heat ($Q_A$), internal heat dissipation from onboard equipment ($Q_e$), and radiative heat loss to deep space ($Q_{sp}$). In thermal equilibrium, the net heat balance becomes zero, and the temperature T, which affects the radiative heat loss term $Q_{sp}$, is uniquely determined.

$$m c_p \frac{dT}{dt} = Q_S + Q_I + Q_A + Q_e - Q_{sp}$$

The thermo-optical properties—solar absorptivity (α) and infrared emissivity (ε)—vary significantly depending on the optical transparency of surface films and the reflectivity of the underlying metal. Even for the same base material, these properties depend on factors such as tape thickness and surface finishing of the metal. Hokkaido University has measurement facilities for α and ε and can obtain accurate values for these properties as needed.

Heat conduction through the hinges connecting the DSAP panels is considered sufficiently small and is therefore neglected in this analysis. Two simplified exposure cases are assumed: one in which a single DSAP panel is exposed directly to outer space, and another in which the main body of the 3U CubeSat without DSAP exposure is considered. Thermal coupling between the DSAP and the main body, as well as shadowing effects caused by the DSAP, are not taken into account.

Figure 7 shows the thermal analysis model and the thermo-optical properties assigned to each surface. In this example, the rear surface of the DSAP panel is covered with aluminum. Three surfaces of the main structure (BODY) correspond to solar array panel surfaces (SAP/AL), while the remaining three surfaces are aluminum. When the main structure is covered with aluminum, incident solar radiation is absorbed by the body, and only a limited amount of heat is dissipated to deep space.

Table 3 summarizes the dimensions, mass, and thermal properties of the BODY and DSAP. The listed material properties correspond to A6061 aluminum alloy and FR4; however, the mass values are represented using equivalent densities that uniformly distribute the masses of mounted components. In this table, C denotes the specific heat and K denotes the thermal conductivity.

Figure 8 shows the arrangement of solar cells on each surface and the corresponding thermo-optical properties (alpha and epsilon) of each material. The specifications of the solar cells are based on the datasheet of the product XTJ Prime. Alodine 1000 is applied to the aluminum surfaces. The polyimide surfaces are fabricated by attaching polyimide tape (P-224) to aluminum treated with Alodine 1000. The thermo-optical properties of the aluminum and polyimide surfaces were measured by Kyushu Institute of Technology (LaSEINE), while those of FR4 were measured using equipment at Hokkaido University. The average thermo-optical properties of the SAP surfaces are calculated based on the area ratios of the constituent materials.

A wide range of simulation cases can be defined based on the external thermal environment, thermo-optical properties, satellite attitude, and solar power generation conditions. The external thermal environments considered in this study are classified into three categories: (1) a sun–eclipse cycle at zero beta angle (cold condition), (2) an average beta angle (typical condition), and (3) a continuous full-sun condition (hot condition). The backside surface of the DSAP panel is modeled using three configurations: (1) aluminum base, (2) aluminum base with polyimide tape, and (3) FR4 base. Three satellite attitude modes are considered: (1) free rotation at 2 deg/s, (2) temporary Earth pointing, and (3) temporary Sun pointing. In addition, two power generation modes are defined: (1) maximum power generation (mission mode and battery charging) and (2) minimum power generation (standby mode). Under maximum power generation, a significant portion of the absorbed solar energy is converted into electrical power, resulting in a noticeable reduction in panel temperature, particularly for the DSAP.

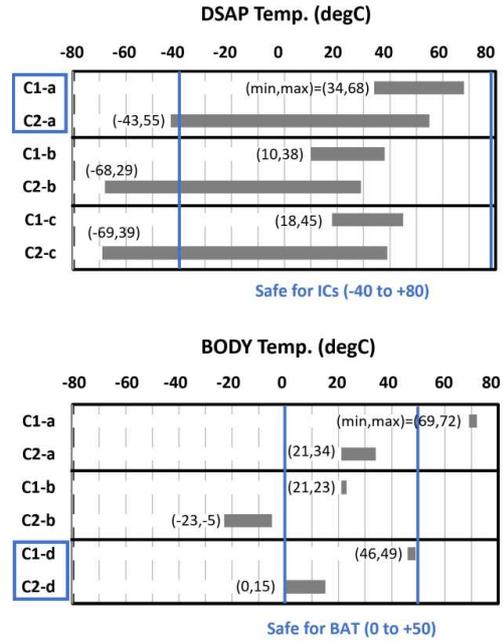

Fig. 9. DSAP and BODY temperature ranges for Case-1 (full-sun, HOT) and Case-2 (β angle = 0, COLD) (both cases assume free rotation at 2 deg/s and minimum power generation)

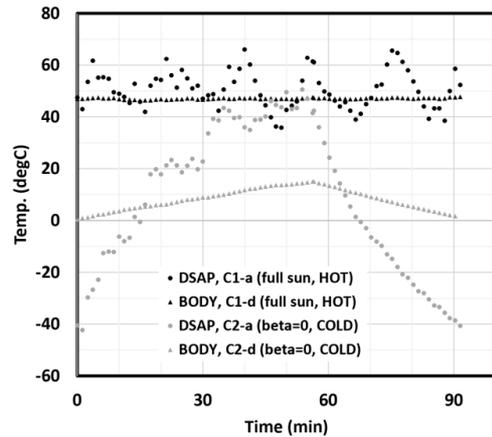

Fig. 10 Analysis results: temperature history over a single orbital period

This paper focuses on two representative cases. Case 1 corresponds to the full-sun hot condition, while Case 2 corresponds to the zero beta angle sun–eclipse cycle cold condition. In both cases, the satellite is assumed to undergo free rotation at 2 deg/s. The assumed values of solar radiation, Earth infrared radiation, and albedo are 1414 W/m², 258

W/m$^2$, and 0.35, respectively, for the hot condition, and 1318 W/m$^2$, 216 W/m$^2$, and 0.25, respectively, for the cold condition. The Earth view factor at an orbital altitude of 400 km is set to 0.885, and the albedo correction factor Ka is assumed to be 0.998.

The maximum power generation of a single solar cell is calculated to be 1.12 W based on a conversion efficiency of 0.307, a solar irradiance of 1353 W/m$^2$, and a cell area of 27 cm$^2$. Under direct solar illumination, a power of 6.7 W can be generated per string consisting of six series-connected cells. However, under free spin rotation, the average generated power is reduced to 1.7 W, corresponding to one quarter of the direct illumination value. The power consumption in standby mode is 2.1 W. Assuming a sunlight duration of 60 minutes and an eclipse duration of 30 minutes per orbit, the required power generation during the sunlight phase is equivalent to 3.2 W, calculated as 2.1 W multiplied by 90 minutes and divided by 60 minutes. Since the satellite has a total of seven parallel strings across the DSAP and BODY, the required power generation per string is 0.46 W. The difference between the maximum power generation of 6.7 W and the minimum required generation of 0.46 W is 6.24 W, which has a strong influence on the temperature of the lightweight DSAP. In both Case 1 and Case 2, the satellite is assumed to operate in the minimum power generation state corresponding to standby mode, with a power generation of 0.46 W per string and an internal heat generation of 2.1 W within the BODY during both sunlight and eclipse phases.

The satellite operates predominantly in a free-rotation mode and temporarily performs Sun-pointing and Earth-pointing maneuvers during mission operations. For example, when the DSAP is oriented toward the Sun under the conditions of Case 1, the DSAP temperature increases significantly, indicating that the timing and duration of such maneuvers must be carefully controlled. Cases 1 and 2 therefore represent the upper and lower bounds of temperature variation driven by environmental factors under minimum power consumption conditions, in which mission operations are inactive.

Figure 9 summarizes the maximum and minimum temperatures obtained over one orbital period. For each case, several subcases are defined by varying the surface configurations of the DSAP and BODY. For the DSAP, subcase (a) corresponds to an aluminum backside, (b) to a polyimide backside, and (c) to an FR4 backside. For the BODY, subcase (a) consists of aluminum or SAP/AL surfaces on all faces except the solar cell areas, subcase (b) consists of polyimide or SAP/polyimide surfaces, and subcase (d) consists of a mixed surface configuration with 70% aluminum and 30% polyimide.

For both Case 1 (C1) and Case 2 (C2), the temperatures of electrical components and the battery must remain within safe operating ranges. The DSAP temperature ranges are 34 to 68 °C for C1-a and −43 to +55 °C for C2-a. Assuming an operational temperature range of −40 to +80 °C for integrated circuit components mounted on the DSAP, the lower limit is slightly exceeded. Nevertheless, configuration (a) is considered superior to the other options. Temperature excursions below −40 °C require further evaluation of component tolerance through thermal vacuum testing.

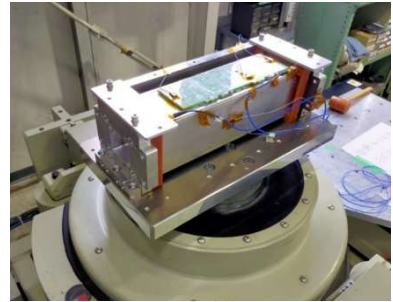

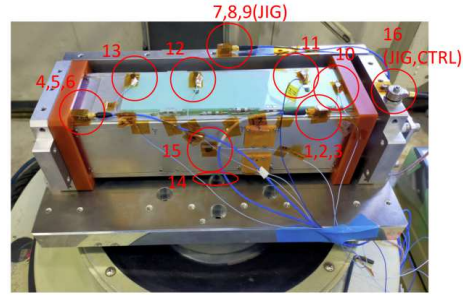

Fig. 11 Apperance of the STM vibration test with acceleration sensor locations

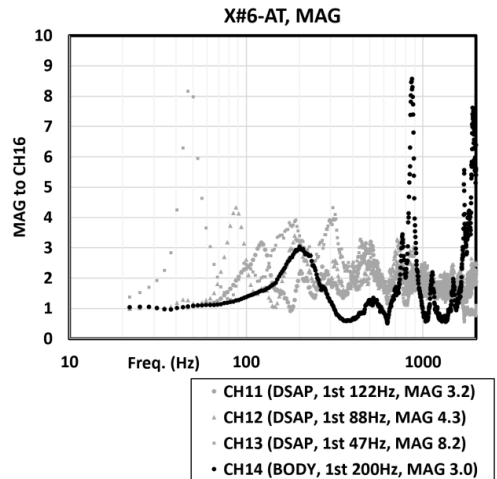

Fig. 12 Response magnification graph (X-axis AT random vibration)

The temperature range of the BODY is 46 to 49 °C for C1-d and 0 to 15 °C for C2-d, which satisfies the recommended battery operating temperature range of 0 to 50 °C.

Figure 10 shows the temperature histories over a single orbital period. Due to its low mass, the DSAP exhibits large temperature fluctuations caused by attitude motion, even under full-sun conditions. In particular, DSAP-C2 shows a large temperature amplitude ranging from −43 to +53 °C. In contrast, the BODY shows almost no temperature variation in Case 1 and only a modest variation from 0 to +15 °C in Case 2.

**6. Vibration Test for the Structural Thermal Model**

Vibration tests of the structural model, equivalent to the Structural Thermal Model (STM), were conducted in June and November 2022. The objectives of these tests were to verify the vibration tolerance of the deployable solar array panels (DSAP) and to validate the newly manufactured vibration test jig. The satellite main body was configured in accordance with the structural analysis model, with internal equipment replaced by stainless steel plates, resulting in a total satellite mass of approximately 4 kg. The satellite was constrained by applying four springs, each providing a force of approximately 47 N, at the rail ends.

Due to facility constraints, a vertical vibration table was used for both tests; however, three-axis excitation using a horizontal vibration table is also feasible. The interface bolt locations between the test jig and the vibration table were positioned 20 cm from the center. Because the test jig extended beyond the 34 cm length of the satellite main body, significant external deflection of the jig was observed. As a result, a large difference in response was measured between the excitation input at the fastening bolts and the response at the upper end of the rails. In the first test, the reference acceleration point was located at the interface bolts. In the second test, the reference point was relocated near the rail end, which resulted in a more stable response waveform.

In both tests, no damage was observed in the mechanical components or FR4 panels of the DSAP, and post-vibration deployment tests were completed successfully. A cover glass (product name C050701, thickness 0.12–0.17 mm), simulating a solar cell, was bonded to the FR4 panel using silicone adhesive, and no cracking was observed. However, each 1.0 mm thick FR4 panel exhibited large flexural deformation during vibration, and contact with a resin jig (brown components visible in the photographs), which simulates the envelope boundary located 6.5 mm from the rail surface, was observed. These results indicate that an increase in panel stiffness is required for subsequent design iterations.

Representative results from the second vibration test are presented below. Figure 11 shows the measurement locations of a total of 16 acceleration sensors. During this test, the shaker was controlled such that the acceleration level at channel CH16 matched the predefined power spectral density (PSD) profile. The test conditions correspond to an overall input level of 4.07 Grms over the frequency range from 20 Hz to 2 kHz. The input PSD levels were set to (0.01, 0.01, 0.0115, 0.0155, 0.0155) $G^2$/Hz at frequencies of (20, 50, 70, 120, 230) Hz, respectively. The response magnification (MAG) relative to the reference point CH16 was calculated from the ratio of the measured PSD values, taking the square root of the PSD ratio.

Figure 12 presents the response magnification spectra. The first resonance peaks of the DSAP components appear at 122 Hz, 88 Hz, and 47 Hz for channels CH11, CH12, and CH13, respectively. For the BODY component, the first resonance peak is observed at 200 Hz at channel CH14, corresponding to 86.2% of the analytically predicted natural frequency of 232 Hz. The measured resonance frequency varies with excitation level, appearing at 291 Hz under the modal survey condition (flat input with an overall level of 0.50 Grms) and at 259 Hz under the 0.5 AT excitation level. All measurement points are required to exceed 60 Hz in the modal survey condition. Channel CH13, which exhibited a resonance frequency of 53 Hz at 0.50 Grms, therefore requires additional stiffness reinforcement. In this test configuration, all DSAP panels were fabricated from FR4; however, the use of aluminum, at least for the outermost portions of the panels, is considered a viable option for increasing stiffness.

## 7. Conclusions

This paper has presented the structural design, structural and thermal analyses (both at launch and after deployment of the solar panels), and STM vibration test results obtained during the initial design phase of the 3U CubeSat HOKUSHIN-1, which is currently under development. Detailed design modifications and additional testing of the deployable solar panels are ongoing. The development of the satellite is expected to be completed within 2023, and on-orbit operations are scheduled to commence after 2024.

### References

1) Shinya Fujita, Ryo Ishimaru, Yuji Sakamoto, Keisuke Maeda, Osamu Okudaira, Yuji Sato, Toshinori Kuwahara, Takafumi Matsui, "Development of Spin Stabilization Control System for the Cosmic Dust Observation CubeSat," *2022 IEEE/SICE International Symposium on System Integration (SII)*, DOI: 10.1109/SII52469.2022.9708909.
2) STMD: Small Spacecraft Technology, https://www.nasa.gov/smallspacecraft (as of May 2023)

**Appendix: Images of the Final Flight Model (FM)**

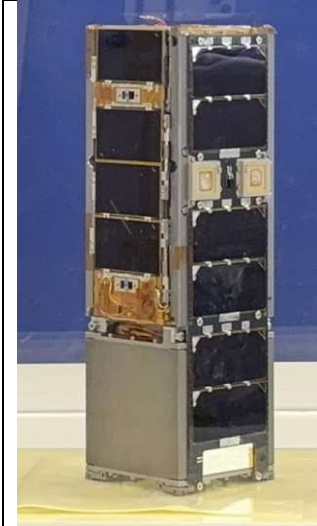

Fig. A-1   -X (left side), +Y (right side) before DSAP deployment (as of February 2025):
The DSAP unit is mounted on the -X face, and two sets of S-band patch antennas are mounted on +Y face.

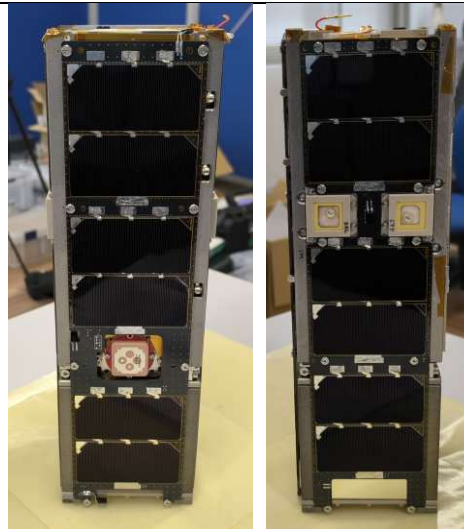

Fig. A-2   +X front (left), -Y front (right) before DSAP deployment (as of February 2025):
A GPS antenna is mounted on +X face, and another two sets of S-band patch antennas are mounted on -Y face.

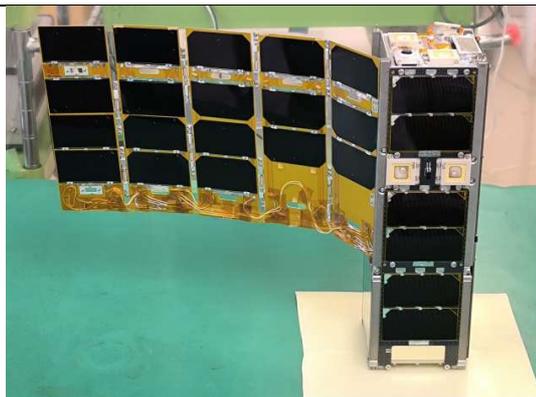

Fig. A-3   Surface of the DSAP after deployment (as of July 2025):
In the final configuration, a total of five deployable panels are mounted on an aluminum plate. At one stage, the deployable panels were temporarily expanded to a 3U size; however, to suppress panel deflection under launch vibration, they were reduced to 2U dimensions. The surface, except for the solar cells, was covered with polyimide flexible printed circuit (FPC).

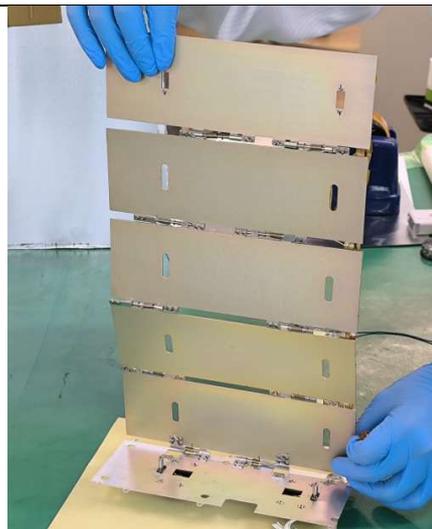

Fig. A-4   Behind the DSAP unit after deployment (as of July 2025):
The rear side of the deployable panels consists solely of aluminum material (treated with Alodine 1200). Considering the surface treatments of both the front and rear sides together, the configuration is close to cases C1-a and C2-a in Fig. 9.   During launch, displacement of the deployable panels is constrained to maintain the envelope by rotating claws that pass through two elongated slots in the panels. The rotating claws are driven to rotate by 90 degrees using a spiral spring, and a shape memory alloy (SMA) actuator is used to provide the driving force to release the latch of the rotation gear.